\documentclass[12pt,preprint]{aastex}

\begin{document}

\shorttitle{Transition Dwarf Oxygen Abundances}
\title{HII Region Oxygen Abundances in Starbursting Transition Dwarf Galaxies}

\author{Kate E. Dellenbusch\altaffilmark{1}\altaffiltext{1}{Visiting Astronomer, Kitt Peak National Observatory,
National Optical Astronomy Observatory, NOAO is operated by the Association of Universities for Research
in Astronomy, Inc. (AURA), under cooperative agreement with the National Science Foundation}}
\affil{Department of Astronomy, University of Wisconsin, 475 N. Charter Street, Madison, WI 53706-1582}
\email{dellenb@astro.wisc.edu}
\author{John S. Gallagher, III}
\affil{Department of Astronomy, University of Wisconsin, 475 N. Charter Street, Madison, WI 53706-1582}
\email{jsg@astro.wisc.edu}
\and
\author{Patricia M. Knezek\altaffilmark{1}}
\affil{WIYN Consortium, Inc., 950 North Cherry Avenue, Tucson, AZ 85726-6732}
\email{knezek@noao.edu}

\begin{abstract}

We present empirical HII region oxygen abundances for a sample of low-luminosity starburst galaxies which are in 
a short lived evolutionary state.  All five galaxies are characterized by centrally concentrated star formation, 
which is embedded in smooth stellar envelopes resembling dE-like systems.  The galaxies also have small gas contents
with typical M$_{HI}$/L$_B$ $\lesssim$ 0.1 resulting in gas exhaustion timescales less than 1 Gyr, even when molecular
gas is considered. We find, compared to other morphologically similar systems, the galaxies of this sample have surprisingly high
oxygen abundances with 12 + log(O/H) $\sim$ 9.0.  We propose that these objects are a subclass of evolved blue compact dwarfs,
which have exhausted most of their gas supply while retaining their metals.  We further propose that we are seeing these objects during
a short phase in which they are nearing the end of their starburst activity, and could become early-type dwarfs. 

\end{abstract}

\section{Introduction}

Although dwarf galaxies are the structurally simplest and most numerous type of galaxy in 
the Universe, the specifics of their formation and evolution remain poorly understood.
One of the main features of dwarf galaxy populations is a grand division between the early-type
spheroidal systems and late-type irregulars; e.g. van den Bergh (1977) emphasized that dwarf
galaxies are typically either very actively forming stars or are inactive, with few dwarfs
between these states.  This raises the question of whether there is an evolutionary link 
between the two main dwarf galaxy families: are there processes which drive late-type dwarfs
to evolve into early-type systems or are these differences ingrained at the time of formation?

The possibility of evolutionary links between early- and late-type dwarf galaxies has been widely
discussed, especially in the context of the large population of dE systems found in clusters of 
galaxies (e.g. Ferguson \& Binggeli 1994; Conselice et al. 2003; Lisker, Grebel \& Binggeli 2006).
In their morphological study Sandage \& Hoffman (1991) proposed that some galaxies in the field
and Virgo cluster should be considered as ``transition dwarfs'' with evolutionary states between
those of the two main dwarf galaxy families.  The Knezek, Sembach, \& Gallagher (1999) 
photometric study of candidate transition dwarfs found that the systems in the Sandage \& Hoffman (1991) sample cover a wide range of
evolutionary states; their status as evolutionary bridges thus remains unclear.  
The situation for the higher luminosity end of the dwarf classification therefore could resemble that for the low luminosity
dSph/dI galaxies, which share traits of both faint dwarf galaxy structural classes (Grebel,  Gallagher \& Harbeck 1993).

In this paper we present HII region oxygen abundances for a sample of moderate luminosity galaxies
which may be a ``missing link" in dwarf galaxy evolution.  These galaxies seem to be evolving via starburts
from a gas-rich, actively star forming state to a gas-poor inactive state.  We use the term ``transition" galaxy in the 
spirit of Sandage \& Hoffman (1991) and Knezek, Sembach, \& Gallagher (1999):  transition dwarf 
galaxies show a mixture of early and late-type characteristics and here, specifically
are experiencing star formation, but with very little HI gas (See Table \ref{obsprop}).  It is not clear, however, 
if these ``transition" dwarfs represent an actual evolutionary transition from Im to dE 
systems, or just one of gas content and level of star formation in the most slowly evolving
part of the dE structural branch.  For example, an alternate model is for these objects to
be highly evolved versions of long lived blue compact dwarfs (BCDs) that eventually may fade to form dE galaxies
(e.g. Bothun et al. 1986; Drinkwater \& Hardy 1991; Marlowe et al. 1992).

Several characteristics of the sample of galaxies presented here indicate that they are in 
phases of rapid evolution.  (1) They are actively forming stars with star formation rates on the order of
a few solar masses per year, but the starburts are fueled by very little HI gas (see Table \ref{obsprop}).  
Star formation can continue in these galaxies
at the current rate for only about 10$^8$ yrs based on HI content.  Although these galaxies may
have significant amounts of molecular gas, it is unlikely to be enough to significantly increase this gas
depletion timescale (e.g. Gordon 1991; Li et al. 1994).  (2) The star formation is currently centrally 
concentrated; much of these galaxies are already stellar fossils.  (3) They have smooth outer 
isophotes, which look more like those from an 
early rather than late-type galaxy (Dellenbusch, Noble, \& Gallagher, in preperation).  
In addition, they are fairly isolated, located in loose group
environments, and show no signs of a recent merger; any dynamical effects are already well in the past.
More details about the structures and star
formation rates of this sample will also be given in Knezek et al. (in preparation).

\section{Observations and Data Reduction}

The data presented in this paper were obtained with the GoldCam spectrograph at the
2.1-m telescope on Kitt Peak, Arizona.  The observations were obtained during six nights
from 2006 January 31 to February 5.  Grating 09 with decker 3 was used.  This yielded
a wavelength range from about 3500 to 7500 \AA\ with a typical FWHM spectral resolution of 8.1 \AA.  
We used the blocking filter WG345 to remove possible overlap between orders.  

Typical exposure times were 30 minutes, except for one night of increased cloud cover,
in which case 20 minute exposures were used.  For all but one galaxy, six 30 minute exposures
were combined for a total integration time of 3 hours.  For NGC 3353, due to 
poor weather conditions and variable cloud cover, only one 20 minute exposure is used in the 
analysis.  The slit was centered on the central HII region of each galaxy.
For NGC 3353 a position angle of 45$^{\circ}$ was used; for the other four galaxies,
the central HII region is sufficiently round that the position angle was left at 90$^{\circ}$.

Spectrophotometric standards from Massey et al. (1988) were observed periodically throughout
each night so flux calibration could be performed.  HeNeAr comparison lamps were observed after
each target exposure to calibrate the wavelength solution.

\begin{figure}
\plotone{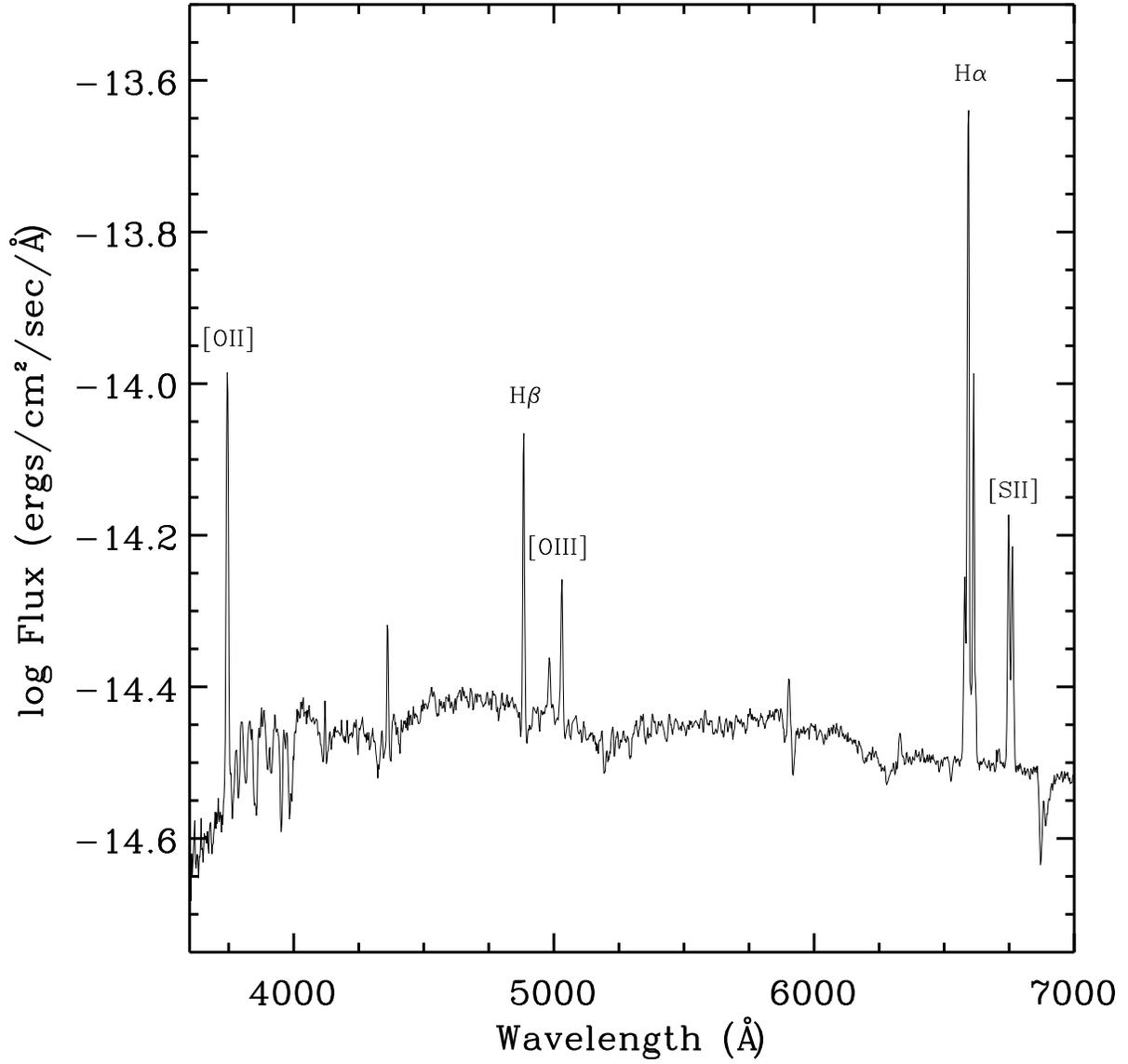}
\caption{Spectrum in log Flux of NGC~3265, one of our sample galaxies.  Key lines are labeled.\label{spectrum}}
\end{figure}

The data were reduced following standard methods in IRAF\footnote{IFAF is distributed by the
National Optical Astronomy Observatories, which are operated by the Association of Universities
for Research in Astronomy, Inc., under cooperative agreement with the National Science
Foundation.}.  The bias level was determined and corrected for each frame using the overscan
region.  Separate zero-second exposures were not used to construct an average bias frame
because of large variations in bias level from exposure to exposure.  Flat-fielding was
accomplished by combining and normalizing a series of quartz-lamp flats.  Once standard CCD
reductions were complete, including the fixing of bad columns, cosmic rays were removed
from the frames using L.A. COSMIC, a rejection routine which utilizes Laplacian edge detection
(van Dokkum, 2001).  The IRAF package DOSLIT was used to calibrate and extract the spectra.  An
example of one of the extracted long exposure spectra is shown in Figure~\ref{spectrum}.

\section{Results and Analysis}

Relative emission-line fluxes were measured from the extracted spectra and corrected for underlying
stellar Balmer absorption and interstellar reddening.  Underlying stellar Balmer 
absorption with an equivalent width of 4 \AA\ was assumed based on the strength of the H$\gamma$ absoprtion
with an approximate correction for the emission line contribution.  Case B recombination with 
T$_e$=10,000 K is adopted along with the SMC extinction curve from Gordon et al. (2003)
to correct for interstellar reddening.  We estimate the errors on flux ratios to be $\sim$ 30$\%$.

Oxygen abundances were estimated indirectly from our corrected HII region spectra using various
abundance diagnostics.  We use the `optimized abundance determination' method described
in Kewley \& Dopita (2002).  This combined method utilizes various indicators to 
cover a range of abundances.  Under this method, the [NII]/[OII] ratio is used to 
estimate abundances in the high metallicity regime with 12 + log (O/H) $\geq$ 8.6.
This estimator is good for all but one of our galaxies (NGC 3353).

The R23 oxygen abundance indicator was also used and produced consistent 
results (see Table \ref{abund}).  This method was suggested by Pagel et al. (1979) and has seen wide use.
The R23 indicator is double valued, having high- and low-abundance branches, and is therefore 
a poor estimator of oxygen abundance near the turnover region between the branches.  This turnover
occurs 8.0 $<$ 12 + log(O/H) $<$ 8.3 (Pilyugin \& Thuan 2005).  On the basis of the strength of [NII] emission,
we estimate abundances for our galaxies
using the upper branch of the R23 indicator from the models of Pilyugin (2000) and Pilyugin \& Thuan 
(2005).  Because NGC 3353 lies near this turnover region, we have a less robust estimate for it's 
oxygen abundance than for our other galaxies.

\section{Discussion}

\begin{figure}
\plotone{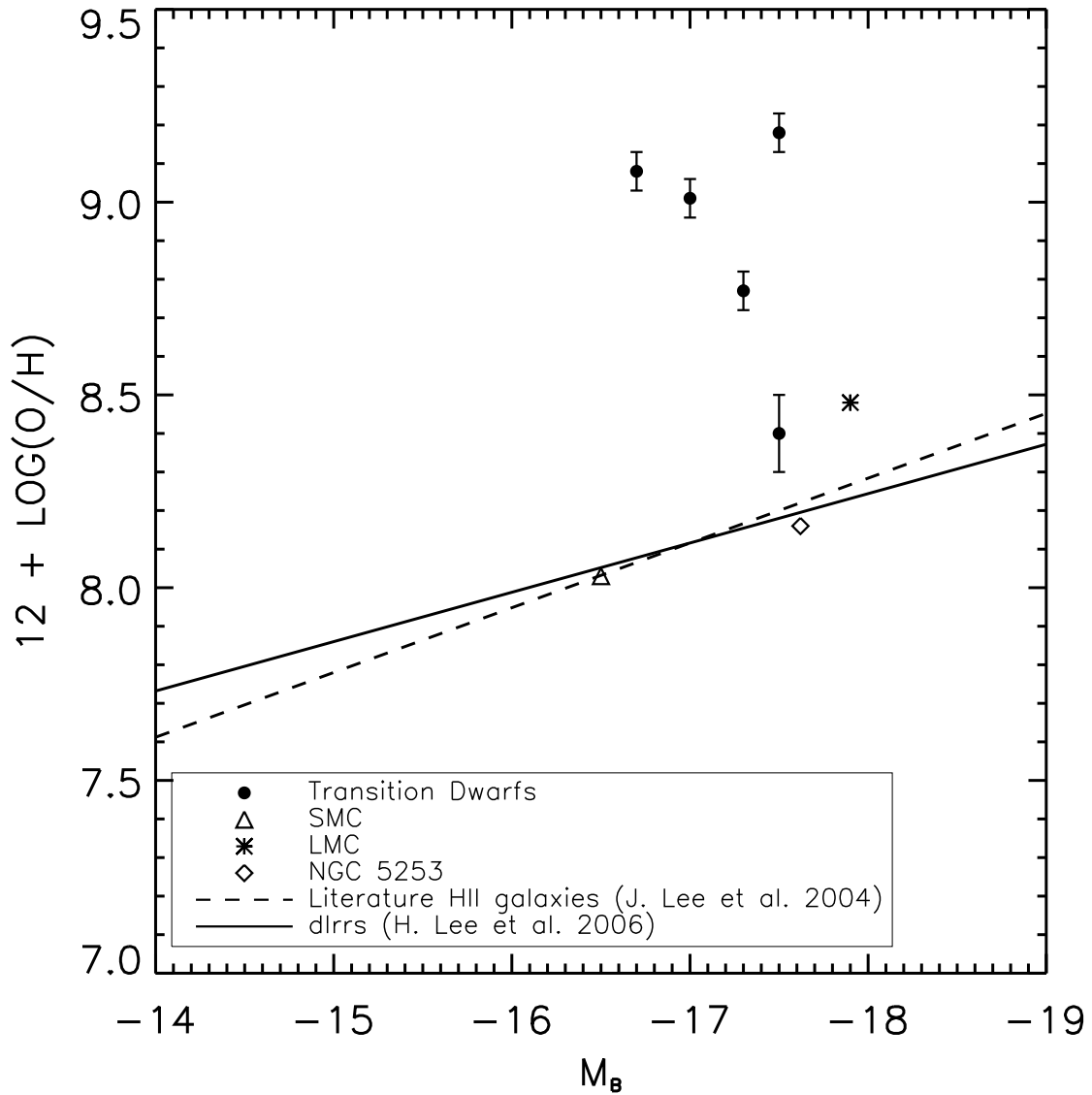}
\caption{Nebular oxygen abundances vs. absolute magnitude in B.  Our sample galaxies are shown by filled circles, with 
	three other dwarfs indicated with symbols as described in the key. For a wider comparison, the dashed line shows
	the best fit for HII galaxies from Lee et al. (2004) and the solid line is the best fit to dIrrs from Lee 
	et al. (2006).  Note, all of our galaxies lie above both of these L-z relations.  We plot [NII]/[OII] derived
	abundances for the transition dwarfs. \label{Lz}}
\end{figure}

The galaxies presented here have surprisingly high HII region oxygen abundances, as compared to 
similar luminosity dI and BCD galaxies with equally strong HII emission lines.  To examine this further, 
we compare these galaxies with other low luminosity
systems by placing them on a luminosity-metallicity plot, shown in Figure \ref{Lz}.
To put these objects in context with other dwarf galaxies, luminosity-metallicity
relations for dIrrs (H. Lee et al. 2006) and HII galaxies (J. Lee et al. 2004) are shown as
solid and dashed lines respectively.  In addition, three other star forming dwarf galaxies are 
also shown; the SMC, LMC, and NGC 5253.  NGC 5253 is included because it is a 
nearby starbursting
dwarf galaxy that exhibits several characteristics similar to the transition dwarfs presented
here, indicating it may be the same class of object.  

The transition dwarfs studied here have high oxygen abundances and lie well above the L-Z$_O$
relations for most other types of small starburst galaxies.  In other words, they are more metal rich than their
similar luminosity HII/BCD/dI counterparts.  It should be noted that there is typically a significant
amount of scatter of about 0.3 dex (e.g. Lee et al. 2006) in the plots from which these mean relations 
are drawn.  Nevertheless, the galaxies
in our sample clearly do not follow the same L-Z$_O$ relation as most other dwarf
galaxies with strong HII emission. This indicates a difference in the past evolution of the galaxies in our sample,
which could be connected to their low gas content and mixed morphologies.  Such high abundances in dwarf elliptical
galaxies is not unprecedented, however. Gu et al. (2006) recently found the dE galaxy IC 225 to have a blue core and
an oxygen abundance of 12 + log(O/H) = 8.98.  We also observed this galaxy and found an oxygen abundance consistent
with their result.

\begin{figure}
\plotone{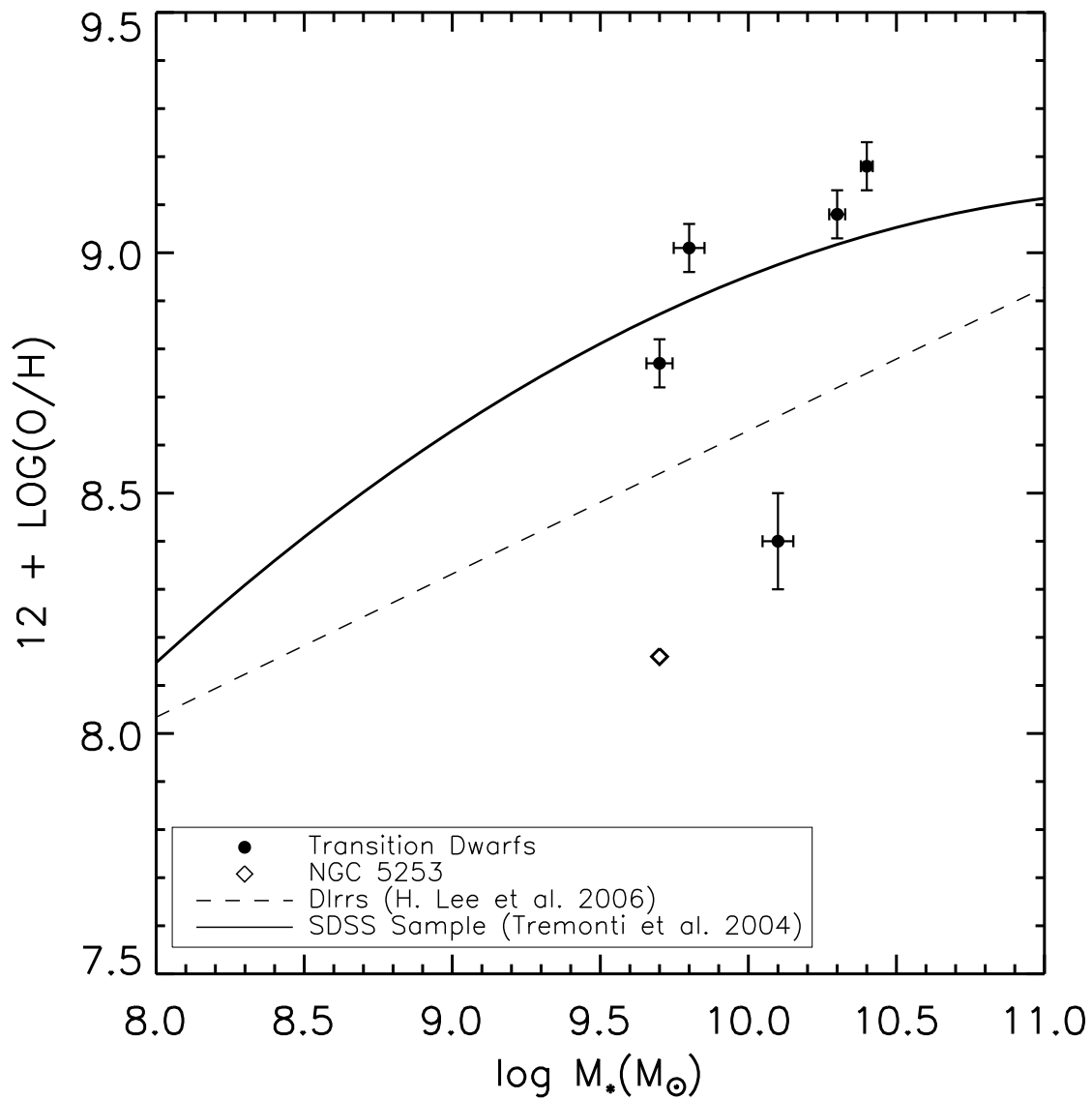}
\caption{Oxygen abundance vs. stellar mass.  Our sample galaxies are shown by filled circles.  The solid line is 
	the fit found by Tremonti et al. (2004) for  over 53,000 star forming SDSS galaxies.  For further comparison,
	the dashed line is an extrapolation of the least-squares fit found by Lee et al. (2006) for dIrrs.
	Our galaxies are consistent with the Tremonti et al. (2004) fit. We plot [NII]/[OII] derived abundances
	for the transition dwarfs. \label{Mz}}
\end{figure}

Because approximate oxygen abundance and luminosity are easily obtained, L-Z$_O$ scaling relations 
have been derived for large samples of galaxies (e.g. Lee et al. 2004; Tremonti et al. 2004).
The underlying physical processes that lead to these relations however, are more closely related
to stellar mass and metallicity than to luminosity.   To examine this more direct physical relation we estimate 
stellar masses from archival (2MASS) K-band magnitudes\footnote{This 
publication makes use of data products from the Two Micron All Sky Survey, which is a joint project 
of the University of Massachusetts and the Infrared Processing and Analysis Center/California 
Institute of Technology, funded by the National Aeronautics and Space Administration and the 
National Science Foundation.}.  Our stellar estimates assume a stellar M/L
ratio calculated from the models of Bell \& De Jong (2001).  We utilize the model coefficients 
for B-R colors, assuming a closed box chemical evolution model and Salpeter IMF.  We find
the galaxies to have log(M$_*$/M$_\odot$) $\sim$ 10.  Metallicity vs. stellar mass is plotted in 
Figure \ref{Mz}.  Although our galaxies again primarily lie above the M$_{\ast}$ - Z$_O$ relation for dwarf
galaxies (dashed line - e.g. Lee et al. 2006), they do scatter around the mean relation found by Tremonti et al. 
(2004) - solid line for star forming SDSS galaxies.  This is explained perhaps in part by the fact that the galaxies
presented here are at the more massive end of the dwarf galaxy spectrum,and therefore in terms of mass, are 
more like the Tremonti et al.(2004) sample than the Lee et al.(2006) sample.  These combined results indicate
that although these systems are unlike other emission line galaxies in L-Z$_O$ space, they are like typical galaxies
of similar stellar mass.

Can we place these objects in an evolutionary context?  An important characteristic that defines these 
galaxies is a dense concentration of ISM at their centers.
It is not obvious how galaxies come to have centrally concentrated ISM, but it is clearly common 
in the Universe as BCD's routinely share this characteristic (e.g. Simpson \& Gottesman 2000; Roye \& Hunter 2000).  
Is there an evolutionary path that leads to this
centralized star formation or are some galaxies born with a centralized gas concentration?  If these
galaxies began with extended gas disks, like a dI, then a trigger would be needed to explain the gas
being accreted into the center (Tajiri \& Kamaya, 2002).  Interactions would provide such a trigger if the gas loses angular
momentum and is therefore able to accrete into the galaxy centers fueling the central starburst.  Alternatively, an
instability associated with low angular momentum could be at work in these galaxies, as suggested by Hoffman et al.
(2003) for Virgo cluster BCDs (see also van Zee et al. 2001).

We postulate that the galaxies presented here are a subclass of low gas-content BCD's.  We further suggest
that we may be seeing BCD's that have retained their metals and are running out of gas and nearing the end of 
their starburst phase.  This is a short lived, as evidenced by the fact that more galaxies of this type are not
observed, but important evolutionary phase.  The end result could be a dE-like system with a metal-rich, younger
stellar core, perhaps related to the objects found in the Virgo Cluster by Lisker et al. (in press).

\acknowledgments

We would like to thank P. Mucciarelli for assisting with the observations.  
This research is supported in part by NSF grant AST 98-03018 and the University of Wisconsin Graduate
School.  JSG also thanks Polichronis Papaderos and Thorsten Lisker for useful discussions of dwarf 
galaxy evolutionary processes.

\clearpage

\begin{deluxetable}{lccc}
\tablewidth{0pt}
\tablecaption{Observed Galaxy Properties \label{obsprop}}
\tablehead{
\colhead{Galaxy} &
\colhead{EW$_{H\beta}$ (\AA)\tablenotemark{a}} &
\colhead{SFR (M$_{\odot}$yr$^{-1}$)} &
\colhead{M$_{HI}$/L$_B$}
}
\startdata
NGC 3265 & 14 & 2.1 & 0.10 \\
NGC 3353 & 23 & 6.0 & 0.30 \\
NGC 3928 & 10 & 1.6  & 0.33 \\
NGC 3773 & 12 & 2.1 & 0.08 \\
IC 745 & 18 & 1.8 & 0.05  \\
\enddata
\tablenotetext{a}{includes 4\AA\ correction for underlying stellar balmer absorption}
\end{deluxetable}

\clearpage

\begin{deluxetable}{lcccc}
\tablewidth{0pt}
\tablecaption{Oxygen Abundance Estimators \label{abund}}
\tablehead{
\colhead{Galaxy} &
\colhead{R$_{23}$} &
\colhead{12 + log(O/H)\tablenotemark{a}} &
\colhead{[NII]/[OII]} &
\colhead{12 + log(O/H)\tablenotemark{b}}
}

\startdata
NGC 3265 &   2.0 & 9.1 & 1.5 & 9.2 \\
NGC 3353 &   8.8 & 8.2 & 0.7 & 8.4 \\
NGC 3928 &   2.0 & 9.1 & 0.8 & 9.1 \\
NGC 3773 &   4.9 & 8.5 & 0.2 & 8.8 \\
IC 745 &     2.0 & 9.1 & 0.5 & 9.0 \\
\enddata
\tablenotetext{a}{oxygen abundance estimated from R$_{23}$}
\tablenotetext{b}{oxygen abundance estimated from [NII]/[OII]}
\end{deluxetable}

\clearpage

\clearpage

\clearpage

\end{document}